\documentclass[pre,twocolumn]{revtex4}
\usepackage{graphicx}
\usepackage{hyperref}
\begin{document}
\title[Gas spreads]{Gas spreads on a heated wall wetted by liquid}
\author{Y. Garrabos}
\author{C. Lecoutre-Chabot}
\affiliation{CNRS-ESEME, Institut de Chimie de la Mati\`{e}re Condens\'{e}e de Bordeaux,\\
Universit\'{e} de Bordeaux I, Avenue du Dr. Schweitzer, F-33608 Pessac Cedex, France}
\author{J. Hegseth}
\affiliation{Department of Physics, University of New Orleans, New Orleans, LA 70148}
\author{V. S. Nikolayev}\email[email:]{vnikolayev@cea.fr}
\author{D. Beysens}
\affiliation{ESEME, Service des Basses Temp\'eratures, CEA-Grenoble, France} \altaffiliation[Mailing
address:] {CEA-ESEME, Institut de Chimie de la Mati\`ere Condens\'{e}e de Bordeaux, 87, Avenue du Dr.
Schweitzer, 33608 Pessac Cedex, France}
\author{J.-P. Delville}
\affiliation{Centre de Physique Mol\'eculaire, Optique et Hertzienne,\\
CNRS, Universit\'e de Bordeaux I, Cours de la Lib\'eration, 33405 Talence Cedex, France}
\date{\today}
\pacs{05.70.Jk, 44.35.+c, 68.03.Cd, 64.60.Fr, 68.35.Rh}

\begin{abstract}
This study deals with a simple pure fluid whose temperature is slightly below its critical temperature and its
density is nearly critical, so that the gas and liquid phases co-exist. Under equilibrium conditions, such a
liquid completely wets the container wall and the gas phase is always separated from the solid by a wetting
film. We report a striking change in the shape of the gas-liquid interface influenced by heating under
weightlessness where the gas phase spreads over a hot solid surface showing an apparent contact angle
\emph{larger} than $90^\circ$. We show that the two-phase fluid is very sensitive to the differential vapor
recoil force and give an explanation that uses this non-equilibrium effect. We also show how these
experiments help to understand the boiling crisis, an important technological problem in high-power boiling
heat exchange.
\end{abstract}\maketitle

\section{Introduction}

Singular properties of a simple fluid \cite{1,2} appear when it is near its critical temperature, $T_c$, and
its critical density, $\rho_c$. When the fluid's temperature, $T$, is slightly lower than $T_c$ and the
average fluid density $\rho $ is close to $\rho_c$ the fluid exhibits perfect wetting (i.e., zero contact
angle) of practically any solid by the liquid phase in equilibrium. In this article we study a system that is
slightly out of equilibrium. Our experiments, performed in weightlessness
\cite{ChimPh,BouldGW,IJT,Sorrento,STAIF} showed that when the system's temperature $T$ is being increased to
$T_c$, the \emph{apparent} contact angle (see Fig.~\ref{zoom} below for definition) becomes very large (up to
$110^\circ$), and the \emph{gas} appears to spread over the solid surface. In section \ref{sec-exp} we
describe our experimental setup that allows the spreading gas to be observed. The gas-liquid interface shape
at equilibrium, which is considered in section \ref{sec-eq}, plays a crucial role as an initial condition for
the gas spreading phenomenon. The sections \ref{sec-c} and \ref{sec-q} deal with the observations of the
spreading gas. A theoretical model that allows this unusual phenomenon to be explained is proposed in section
\ref{sec-t}. In the section \ref{sec-b}, we discuss the boiling crisis, a phenomenon that plays an important
role in industrial applications, and how it is relevant to the spreading gas.

\section{Experimental setup}\label{sec-exp}

We report results that were obtained and repeated using several samples of ${\rm SF}_{6}$ ($T_c=318.717$~K,
$\rho_c=742$~kg/m$^3$). These samples were heated at various rates in cylindrical cells of various aspect
ratios on several French/Russian and French/American missions on the Mir space station using the Alice-II
instrument \cite{3}. This instrument is specially designed to obtain high precision temperature control
(stability of $\approx 15\,\mu$K over 50 hours, repeatability of $\approx 50\,\mu$K over 7 days). To place
the samples near the critical point, constant mass cells are prepared with a high precision density, to
0.02\%, by observing the volume fraction change of the cells as a function of temperature on the ground
\cite{4}.

\begin{figure}[hbt]
  \begin{center}
  \includegraphics[height=5cm]{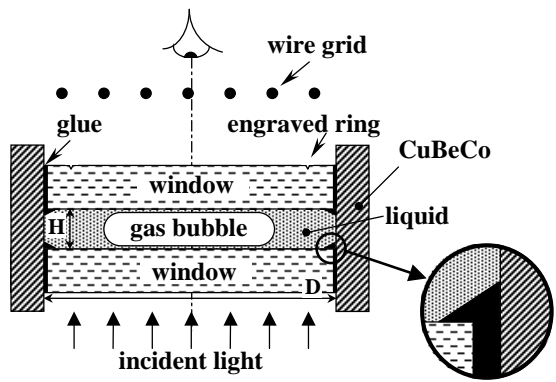}
  \end{center}
\caption{Sketch of a cross-section of the cylindrical sample cell (with parallel windows).  The fluid volume
is contained between two sapphire windows that are glued to a CuBeCo alloy ring. The dimensions $H$ (see Table
\ref{tab1}) and $D$(=12 mm) of the cell are indicated. Some glue is squeezed into the cell. The thickness of
the glue layer is exaggerated for illustration purposes. In weightlessness, the gas bubble should be located
in the middle of such an 'ideal' cell, see sec.~\ref{sec-eq} for the discussion.}\label{fig1}
\end{figure}

A fluid layer was sandwiched between two parallel sapphire windows and surrounded by a copper alloy housing
in the cylindrical optical cell, the axial section of which is shown in Figure~\ref{fig1}.
We consider here three cells of the same diameter $D=12$~mm, the other parameters of which are shown in
Table~\ref{tab1}. The liquid-gas interface was visualized through light transmission normal to the windows.
Since the windows were glued to the copper alloy wall, some of the glue is squeezed inside the cell as shown
in Fig.~\ref{fig1}. This glue forms a ring that blocks the light transmission in a thin layer of the fluid
adjacent to the copper wall making it inaccessible for observations. Because of this glue layer, the windows
may also be slightly tilted with respect to each other as discussed in sec.~\ref{sec-eq}.

A 10~mm diameter ring was engraved on one of the windows of each cell in order to calibrate the
size of the visible area of the cell images as can be seen in each
image. An out-of-focus wire
grid, designed to visualize \cite{Gulf} fluid inhomogeneities and/or a fluid
flow through light refraction, was also used. The grid was occasionally moved out of the light path, so that it is not always present in all images.
\begin{table}
\begin{ruledtabular}
\begin{tabular}{c|c|c}
  Cell number & Cell thickness $H$ (mm) & $(\rho-\rho_c)/\rho_c$ (\%) \\
  8 & 3.016 & 0.85 \\
  10 & 1.664 & 0.25 \\
  11 & 4.340 & 0.87 \\
\end{tabular}
\end{ruledtabular}
\caption{Physical parameters of the experimental cells. Cell 11 has a movable piston to change the cell
volume. However, the volume was kept constant during these experiments.}\label{tab1}
\end{table}

The sample cell is placed inside of a copper Sample Cell Unit (SCU) that, in turn, is placed inside of a
thermostat. Heat is pumped into and out of the SCU using Peltier elements and heaters. The
temperature is sampled every second and is resolved to $1\mu$K.

Similar ground based experiments were done before these experiments using a copy of the same instrument. The
gravity forces push the denser liquid phase to the bottom of the cell and completely different behavior is
seen, see \cite{5}.

\section{Bubble position at equilibrium under weightlessness}\label{sec-eq}

The gas volume fraction $\phi$ (volume of the gas divided by the total cell volume) is defined by $\rho$ and
the densities of gas and liquid for the given temperature. In our experiments, $\phi\approx 0.5$ and the gas
bubble is flattened between the windows (Fig.~\ref{fig1}) due to the large aspect ratio $D/H$ of the cell.

\begin{figure*}[htb]
  \begin{center}
  \includegraphics[height=5cm]{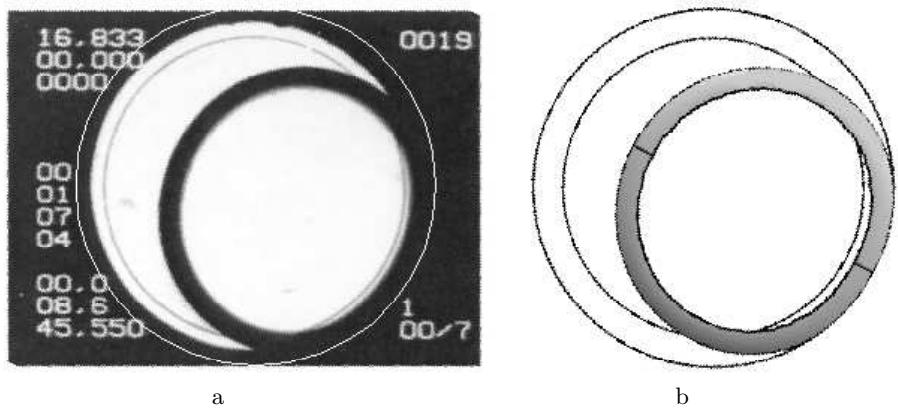}\\
  a\hspace*{6cm}b
  \end{center}
\caption{The experimental image of cell 10 at room temperature (a) and the equilibrium bubble shape simulated
for the tilt angle of $0.46^\circ$ (b). When superposed, the images (a) and (b) give almost perfect match. The
outer white circle in (a) (black in (b)) shows the actual location of the cell wall. The inner black circles
in (a) and (b) correspond to the engraved ring that allows the superposition to be made. The dark space
between two these circles in the image (a) is made by the ring of glue as shown in Fig.~\ref{fig1}. The image
(b) is a frontal projection of the bubble shown in Fig.~\ref{Theor-inclined}. }\label{Theor-Exp}
\end{figure*}

Let us first consider an ideally cylindrical cell as opposed to the real cell. At equilibrium, the windows and
the copper wall are wetted by the liquid phase. Because the van der Waals forces from the walls act to make
the wetting film as thick as possible, the weightless bubble should be located in the cell's center. Because
the bubble is flattened and occupies one-half of the available volume, the distance of such a centered bubble
to the copper wall is large (Fig.~\ref{fig1}). The lateral centering forces are then much weaker than the
centering forces in the direction of the cell axis. Any small external influences in the real cell can
displace the bubble laterally from the cell's center. This displacement is illustrated in
Fig.~\ref{Theor-Exp} that shows cell 10 at room temperature.
We note that there are two kinds of external influences that are easily identified: residual accelerations in
the spacecraft and cell asymmetry.

Bubble images for cells 8 and 10 were recorded in four Mir missions between 1996 and 2000. Several images are
reported in \cite{ChimPh} (Cassiopeia mission, 1996) and in Fig.~\ref{GMSF2} below (GMSF2 mission, 1999) for
cell 10. It is extremely likely that the space station changed its position with respect to the residual
gravity vector between these runs. The bubble position with respect to the cell, however, always remained the
same. The bubble location also varies from cell to cell without any dependence on the station's orientation.
Therefore, we have no reason to attribute the off-center position of the bubble to the residual gravity.

\begin{figure}[bht]
  \begin{center}
  \includegraphics[height=4.5cm]{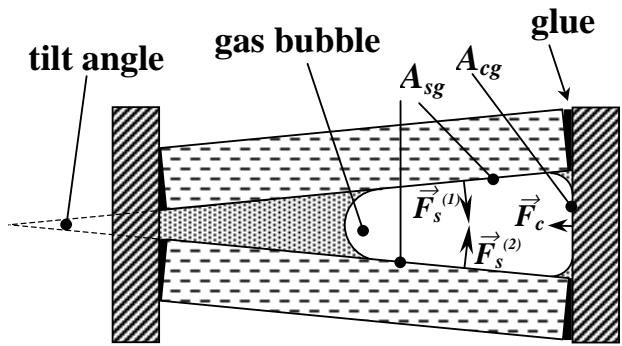}
  \end{center}
\caption{Sketch of a cross-section of the sample cell with the tilted windows at
 equilibrium in weightlessness. The wetting film is not shown. The window tilt is possible due to the
existence of a space between the window's edge and the copper wall, which is filled by glue. This space and
the tilt are exaggerated for illustration purposes. Based on the manufacturing process a maximum tilt angle
of $\approx 1^\circ$ is possible. The glue squeezed into the cell is not shown. The reaction forces that act
on the gas bubble are shown with arrows. The contact areas of the gas bubble with the solid are
indicated.}\label{tilt}
\end{figure}

Although the cells were manufactured with high precision, the cell windows could
not be exactly parallel
because of the glue layer as shown in Fig.~\ref{tilt}. In the rest of this section we will discuss the influence of
the windows' tilt on the position and on the shape of the bubble.

When the bubble's surface is curved, there is a constant excess pressure $\Delta
p$ inside the bubble defined by the Laplace formula
\begin{equation}\label{Lap}
\Delta p=\sigma K,
\end{equation}
where $\sigma$ is a surface tension and $K$ is the surface curvature. This excess pressure acts on all parts
of the bubble interface. In particular, it acts on the part $A_{sg}$ (where the index $s$ stands for
``sapphire" and $g$ for ``gas") of the flat window surface that contacts the gas directly (or, more
accurately, through a wetting film that we assume to be of homogeneous thickness). This pressure creates
reactions forces $\vec{F}_s^{(1)}$ and $\vec{F}_s^{(2)}$ at each window, that act on the bubble. Each of
these forces is perpendicular to the corresponding window. The absolute values of $\vec{F}_s^{(1)}$
 and $\vec{F}_s^{(2)}$ are
equal to $A_{sg}\Delta p$. When the windows are exactly parallel, $\vec{F}_s^{(1
)}+\vec{F}_s^{(2)}=0$ and the
bubble remains at the cell's center. When the windows are tilted with respect to each other, the non-zero force
$\vec{F}_s^{(1)}+\vec{F}_s^{(2)}$ pushes the bubble in the direction of the increasing cell thickness. This
motion continues until the bubble touches (through a wetting film) the copper wall of the cell, thus forming
a contact spot of the area $A_{cg}$, where the index $c$ stands for ``copper".
This direct contact with the
solid results in another reaction force $\vec{F}_c$ with the absolute value
 $A_{cg}\Delta p$, such that
\begin{equation}\label{equil}
\vec{F}_s^{(1)}+\vec{F}_s^{(2)}+\vec{F}_c=0
\end{equation}
in equilibrium, see Fig.~\ref{tilt}.

There are two equivalent ways to find the bubble shape at equilibrium. One can solve Eq.~(\ref{Lap}) that
reduces to $K=$const, where the constant is obtained from the condition of the given bubble volume. The
bubble volume is defined by the known gas volume fraction and the cell volume. One can also minimize the
gas-liquid interface area with a bubble volume constraint. In both cases boundary conditions must be
satisfied (zero contact angle in our case). The resulting bubble shape obviously depends on the cell
geometry. It is also nearly independent of temperature as can be seen from Eq.~(\ref{equil}), because all
three terms of this equation are proportional to the surface tension $\sigma$, so that this force balance
remains valid even near $T_c$, where $\sigma$ disappears. There are, however, several sources of weak
temperature dependence of the bubble shape. First, there is weak dependence of the gas volume fraction $\phi$
on temperature at constant average density $\rho$. This small deviation is smallest at the critical density
$\rho_c$ and slightly greater in these experiments due to the very small deviation (see Table~\ref{tab1}) of
$\rho$ from $\rho_c$. Second, the curvature $K$ depends on the thickness of the wetting film that increases
near $T_c$. The wetting film remains small, however, in comparison with the cell thickness. Both of these
effects are very weak.

The force $\vec{F}_c$, which is directed horizontally in Fig.~\ref{tilt}, causes a distortion of the bubble.
This distortion results in an oval image in Fig.~\ref{Theor-Exp} instead of a circle. The degree of
distortion increases with the tilt angle because so does $\vec{F}_c$. This distortion can thus be used to
estimate the tilt angle.

For these constant volume gas bubbles, the degree of distortion should decrease with increasing cell
thickness $H$ for the same window tilt. A larger value of $H$ results in a less compressed (more sphere-like)
bubble shape with less area in contact with the wall. This smaller bubble curvature results in a smaller
value for $\Delta p$ according to Eq.~(\ref{Lap}). Consequently, the force $\vec{F}_ c$, the area $A_{cg}$ of
the contact with the copper wall, and the bubble distortion are smaller. This window tilt hypothesis is
consistent with observations: we were not able to detect any distortion of the gas bubble in cell 8 (see
Fig.~\ref{Pegase4608}a below that corresponds to the nearly equilibrium shape) that is approximately twice as
thick (Table~\ref{tab1}) as cell 10 shown in Fig.~\ref{Theor-Exp}. There is, however, some tilt in cell 8
because the bubble touches the wall. We expect that the tilt angles in all of the cells are of the same order
of magnitude because they were all manufactured using the same method.

\begin{figure}[htb]
  \begin{center}
  \includegraphics[height=5cm]{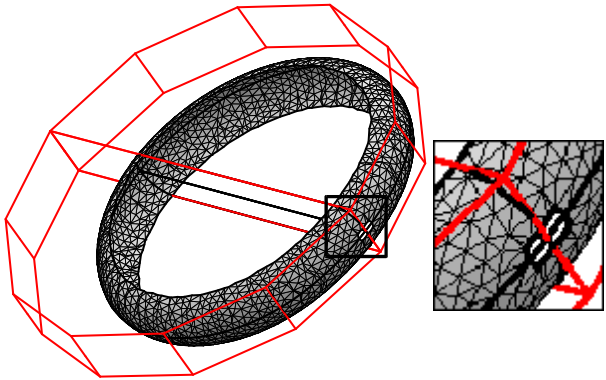}
  \end{center}
\caption{ The result of a 3D finite element calculation of the equilibrium gas-liquid interface for cell 10
with a window tilt angle of $0.46^\circ$. The vertices of the polygonal lines indicate the location of the
cylindrical copper wall and they are shown to guide the eye. A shape of the circular cylinder was input to
the simulation. The contact angle is zero. A part of the image marked by the square is enlarged to show the
contact area $A_{cg}$ of the gas with the copper wall (a small white rectangle crossed by two symmetry
lines). The contact areas $A_{sg}$ with the windows have the oval shape. The projection of this bubble shape
to the cell window is shown in Fig.~\ref{Theor-Exp}b.}\label{Theor-inclined}
\end{figure}

To verify the window tilt hypothesis, we performed a 3D numerical simulation of the bubble surface by using the
Surface Evolver finite element software \cite{Evolver}. The result of this
calculation is shown in
Fig.~\ref{Theor-inclined} for cell 10.
The experimentally observed bubble deformation matches the calculation performed for a tilt angle of
$0.46^\circ$, see Fig.~\ref{Theor-Exp}. The simulation resulted in the interface curvature
$K=1.389$~mm$^{-1}$ and in $A_{cg}=0.150$~mm$^2$ calculated for the bubble volume $V\phi=26.675$~mm$^3$. From
this data, it is easy to calculate the effective acceleration $g_{eff}$ that would create the equivalent
buoyancy force $F_c=(\rho_L-\rho_V)V\phi\,g_{eff}=K\sigma\,A_{cg}$. It turns out that $g_{eff}=1.55\cdot
10^{-3}\,g$ for $T=290$~K, where $g$ is the gravity acceleration on Earth. This $g_{eff}$ acceleration is much
larger than the residual steady accelerations in the Mir space station ($\sim 10^{-6}g$) and this shows that
the observed bubble deformation is not caused by residual accelerations. We conclude that the window tilt
hypothesis about the origin of the bubble deformation and its off-center position is correct.

A similar off-center bubble position was observed under weightlessness in a cell similar to ours by Ikier
\textit{et. al.} \cite{14} and was attributed to a residual acceleration. However, they report only one run
in a single cell making the actual cause of the bubble off-centered position impossible to verify.

\section{Continuous heating experiments}\label{sec-c}

\begin{figure}[htb]
  \begin{center}
  \includegraphics[height=6cm]{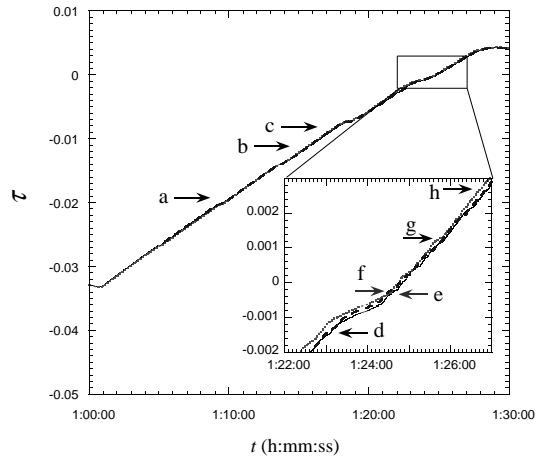}
  \end{center}
\caption{Reduced temperature evolution for the image sequences shown in Fig.~\ref{GMSF2}
 (solid line),
Fig.~\ref{GMSF3} (dotted line), and Fig.~\ref{PrePersU2} (dashed line). The temperature values that
correspond to each of the images (a-h) shown in these figures are indicated by arrows and the corresponding
letters. The definition of $\tau$ is discussed in the text. The temperature is measured in the body of the
SCU. The vicinity of the critical point is enlarged in the insert.}\label{temp}
\end{figure}

In the continuous heating experiments, the cells 8, 10 and 11 were heated
nearly linearly in time $t$. The evolution of the non-dimensional temperature
$\tau$ for each of these experiments is shown in Fig.~\ref{temp}.
The parameter $\tau$ is defined as $(T-T_{coex})/T_c$, where $T_{coex}$ is the temperature of the coexistence
curve that corresponds to the fluid's average density shown in Table~\ref{tab1}. Note that $T_{coex}$ differs
from $T_c$ only by $1-50\,\mu$K because the density is very close to $\rho_c$ for all cells. A 40~min
temperature equilibration at $\tau\approx-0.033$ preceded the heating. The mean value of ${\rm d}T/{\rm d}t$
at $T_c$ was $\approx 7.2$~mK/s.

\begin{figure*}[htb]
  \begin{center}
  \includegraphics[height=5cm]{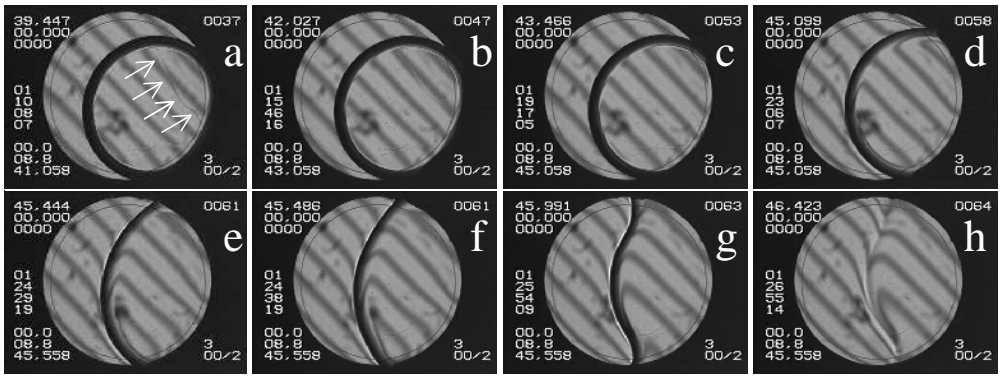}
  \end{center}
\caption{Time sequence of images of the cell 10 during the continuous heating through the critical point. The
temperature values that correspond to each of the images (a-h) shown in these Figures are indicated in
Fig.~\ref{temp} by arrows and the corresponding letters. This run is a repeat of the run shown in Fig.~2 of
\protect\cite{ChimPh}. The gradual increase of the apparent contact angle as the gas spreads with increasing
temperature is clearly seen. The time corresponding to each image is shown to the left of the cell in the
middle. The magnified upper regions close to the contact line from the images (e-g) are shown in
Fig.~\ref{zoom}.}\label{GMSF2}
\end{figure*}

Figure~\ref{GMSF2} shows the time sequence of the images of the cell 10.
\begin{figure*}[htb]
  \begin{center}
  \includegraphics[height=5cm]{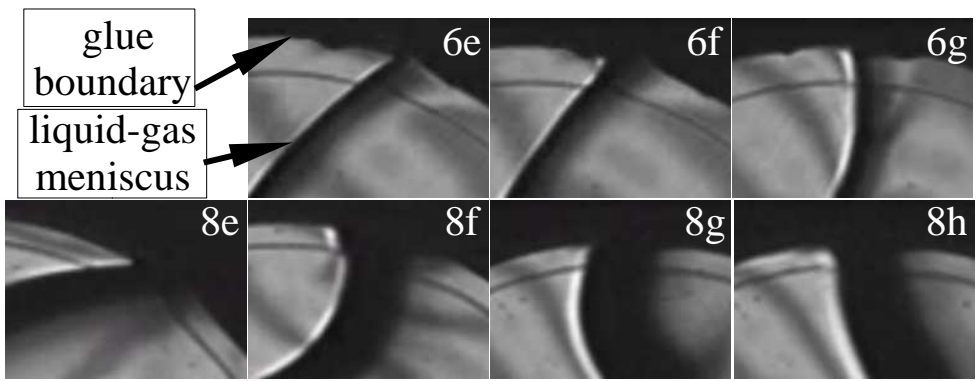}
  \end{center}
\caption{The magnified upper regions close to the contact line from the images Fig.~\ref{GMSF2}(e-g) and from
the images Fig.~\ref{GMSF3}(e-h). The apparent contact angle can be ``measured" as the angle between the
tangents to the black glue boundary and the liquid-gas meniscus. The latter corresponds to the boundary
between the wide dark and narrow bright stripes on the images. The liquid domain is to the left from the
meniscus. One can see that this apparent contact angle exceeds $90^\circ$ in the images \ref{GMSF2}g and
\ref{GMSF3}f.}\label{zoom}
\end{figure*}
The interface appears dark because the liquid-gas meniscus refracts the normally incident light away from the
cell axis. After the temperature ramp was started but still far from the critical temperature, the bubble
shape changed. The contact area $A_{cg}$ of the gas with the copper wall appears to increase. In other
systems the wetting film under a growing vapor bubble is observed to evaporate \cite{Tong}. In near-critical
fluids, however, the heat transfer processes are more complex \cite{Straub}. In this system we believe that
there may be a similar drying process, i.e., at some time the thin wetting film that separates the gas from
the copper wall evaporates. In fact, we have observed low contrast lines that appear within the $A_{sg}$ area
when the heating begins. An example of such a line is indicated in Figure~\ref{GMSF2}a by the white arrows.
The out-of-focus grid shows that these lines correspond to a sharp change in the wetting film thickness. These
lines are most likely triple contact lines and we have actually seen them pinned by an imperfection on the
windows as they advance and retreat in other experiments. Since the heat conductivity of copper is larger
than that of sapphire, the heat is supplied to the cell mainly through the hotter copper wall. Therefore the
film should evaporate on the copper wall even earlier than on the sapphire. A more refined analysis of the
contact line motion will be discussed elsewhere.

The increase of the $A_{cg}$ area is accompanied by an evident increase in the apparent contact angle, see
Fig.~\ref{GMSF2}d--f and the corresponding magnified images in Fig.~\ref{zoom}. Near the critical temperature
the apparent contact angle becomes \emph{larger} than $90^\circ$! We will analyze these effects theoretically
in section~\ref{sec-t}.

While crossing the critical point, the vapor bubble is rapidly evolving. At
$T\approx T_c$, the surface tension vanishes, the bubble's relaxation from
surface tension is negligible, so that the interface shape is defined by the
variation of the local evaporation rate along the interface. The evaporation is
stronger at the parts of the interface closest to the copper heating wall.
This effect leads to the waved interface shape shown in Fig.~\ref{GMSF2}g.
Diffusion causes the disappearance of the interface at $T>T_c$ as shown in
Fig.~\ref{GMSF2}h.

\begin{figure*}[hbt]
  \begin{center}
  \includegraphics[height=5cm]{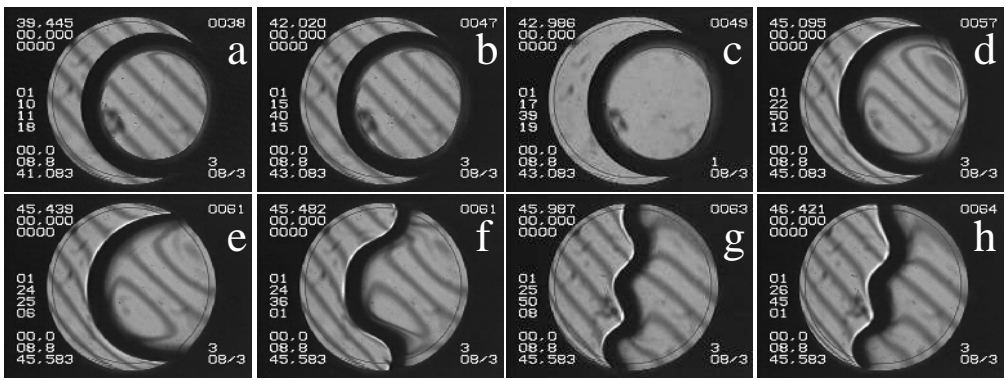}
  \end{center}
\caption{Time sequence of the images from cell 8 during continuous heating through the critical point. Images
(a-h) were taken exactly for the same values of temperature (shown in Fig.~\ref{temp}) as corresponding
images in Fig.~\ref{GMSF2}. The magnified upper regions close to the contact line from the images (e-h) are
shown in Fig.~\ref{zoom}.}\label{GMSF3}
\end{figure*}

Figure~\ref{GMSF3} shows the time sequence of the images of the cell 8, which
is approximately twice as thick as cell 10.
The images in Fig.~\ref{GMSF3} were taken exactly for the same values of the non-dimensional temperature
$\tau$ (shown in Fig.~\ref{temp}) as the corresponding images in Fig.~\ref{GMSF2}. The force $\vec{F}_c$
pushes the bubble against the cell wall as in the case of cell 10. As discussed above, this force is weaker
than for the cell 10 because the bubble appears almost circular at equilibrium (see Fig.~\ref{GMSF2}a and
Fig.~\ref{GMSF3}a). By comparing images (e) of both sequences, we can also see that the vapor spreads slower
in cell 8. The increase of the apparent contact angle is also slower. The waved shape interface appears
earlier in Fig.~\ref{GMSF3}f, i.e. farther from $T_c$ than for the cell 10. The interface is still quite
sharp in Fig.~\ref{GMSF3}h, while it has already diffused in the case of the thinner cell
(Fig.~\ref{GMSF2}h). This difference can be explained by the difference in the liquid-gas interface area,
which is roughly proportional to the cell thickness. The surface tension force that tends to maintain the
convex shape is not as strong for the thicker cell where a larger fluid volume has to be moved during the
same time. The diffusion time is larger for cell 8 because the size of the inhomogeneity (i.e. interface) is
larger.

\begin{figure}[htb]
  \begin{center}
  \includegraphics[height=5cm]{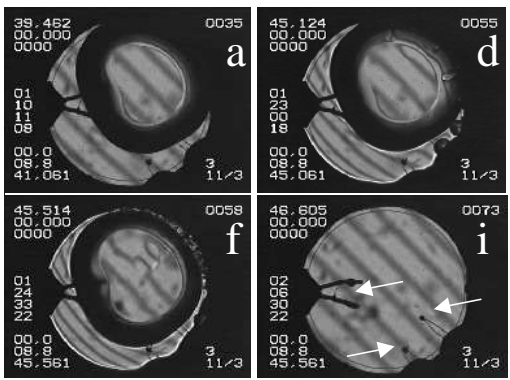}
  \end{center}
\caption{Time sequence of images of the cell 11 during the continuous heating through the critical point. No
bubble spreading is seen. The bubble does not touch the copper wall. The images
(a, d, f) were taken exactly
for the same values of temperature (shown in Fig.~\ref{temp}) as corresponding images in Fig.~\ref{GMSF2} and
Fig.~\ref{GMSF3}. The cell 11 contains three thermistors shown in image (i) by arrows. This image was taken
after the temperature equilibration above $T_c$.}\label{PrePersU2}
\end{figure}

Figure~\ref{PrePersU2} shows the time sequence of the images from cell 11, which is thicker than both the
cells 8 and 10.
The images (a), (d) and (f) were taken for the same values of non-dimensional temperature as corresponding
images in Figs.~\ref{GMSF2} and \ref{GMSF3}. This cell contained three wetted thermistors
(Fig.~\ref{PrePersU2}i) that constrain the bubble surface (Fig.~\ref{PrePersU2}a). The bubble is only slightly
squeezed by the windows so that the reaction forces that act on the bubble at equilibrium are weak. As a
result, the bubble does not touch the copper heating wall at all. Although this is not clear in the image
(a), because of the glue near the copper wall, it is clear in image (f) where small newly formed bubbles
separate the initial bubble from the wall. These bubbles form from the local overheating of the fluid between
the large bubble and the copper wall. There is enough fluid between the large bubble and the wall so that a
small bubble may grow in it. These small bubbles push the large bubble away from the wall before any
coalescence can take place.

The comparison of these three experiments clearly shows that in order to obtain the bubble spreading, the
bubble needs to have a direct contact with the heating wall, i.e. to be pushed to the heating wall by some
force. Note that none of the images show any evidence of steady fluid motion that would be necessary to
maintain the distorted bubble shapes in Figs.~\ref{GMSF2} and \ref{GMSF3}. We conclude that this distortion
of the bubble equilibrium shape cannot be caused by fluid motion.

Similar continuous heating experiments are reported by Ikier \textit{et. al.} in \cite{14}. However, a smaller
heating rate (1.7~mK/s) and erratic accelerations of the cell did not allow gas spreading to be observed.

\section{Quenching experiments}\label{sec-q}

\begin{figure*}[htb]
  \begin{center}
  \includegraphics[height=5cm]{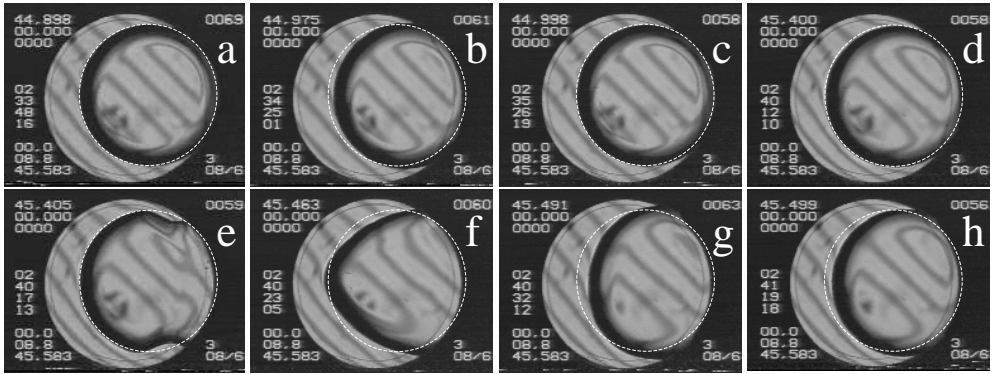}
  \end{center}
\caption{Time sequence of images of cell 8 during two 100~mK quenches. The gas spreads during each quench
that lasts about 12~s. The equilibrium position of the vapor bubble with respect
 to the cell is shown by the
white circle in each image for comparison. }\label{Pegase4608}
\end{figure*}

\begin{figure}[htb]
  \begin{center}
  \includegraphics[height=6cm]{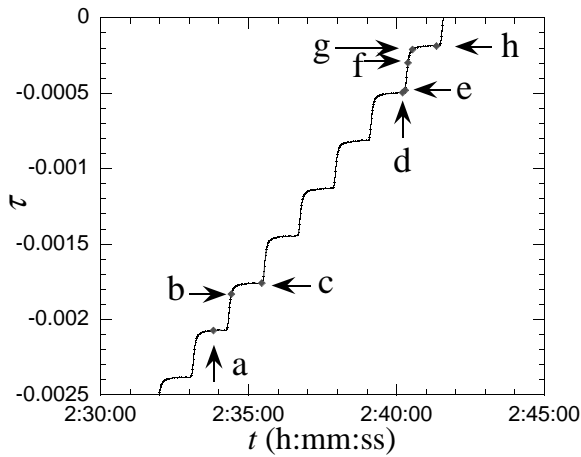}
  \end{center}
\caption{Temperature evolution during the series of quenches. The points that correspond to each of the
images in Fig.~\ref{Pegase4608}(a-h) are indicated by arrows and corresponding letters. The temperature is
measured in the body of the SCU.}\label{Pegase4608temp}
\end{figure}

Figure~\ref{Pegase4608} shows the time sequence of the images of cell 8
when it was heated by 100~mK quenches as shown in Fig.~\ref{Pegase4608temp}.
While the heating rate is quite large during each quench, the time average of the heating rate 1.4~mK/s is
smaller than that during the continuous heating due to the waiting time of $\approx 60$~s after each quench.
During this waiting time a partial equilibration takes place. The images (a--c) show a slight bubble
spreading that appears during a quench that is farther from the critical point than the quench shown in
images (d-h). After each quench as soon as the heating stops, the bubble interface begins to return to its
initial form (Fig.~\ref{Pegase4608}c,d). This shows that the spreading vapor is caused by a non-equilibrium
effect. The second quench that precedes the crossing of the critical point (Fig.~\ref{Pegase4608}d--h) shows
very rapid interface motion accompanied by fluid flows.

While the interface returns to its initial state during the waiting time of the first quench
(Fig.~\ref{Pegase4608}c), it does not return in the second quench (Fig.~\ref{Pegase4608}h). This occurs
because the characteristic equilibration time grows dramatically near $T_c$.

The same phenomenon of spreading gas was also observed during the heating of CO$_2$ cells in other experiments
(Pegasus BV4705, Post-Perseus F14) carried out by our group in the Mir station. However, these experiments
were not designed to study the spreading gas and we do not discuss them here.

\section{Interface evolution during the heating}\label{sec-t}

The above experimental data showed that the spreading gas and the associated interface deformation are caused
by an out-of-equilibrium phenomenon. This is especially demonstrated by the analysis of the interface shape at
equilibrium (sec.~\ref{sec-eq}) and by the return to the equilibrium shape after each quench in
sec.~\ref{sec-q}. In this section we analyze possible causes of the spreading gas. Two causes are considered:
Marangoni convection due to the temperature change $\delta T_i$ along the gas-liquid interface and the
differential vapor recoil.

\subsection{Marangoni convection}

If a temperature change $\delta T_i$ exists, it will create a surface tension change $\delta\sigma =( {\rm
d}\sigma/{\rm d}T)\delta T_i$ that will drive a thermo-capillary (Marangoni) flow in the bulk of both fluids
\cite{7,8,9}. The images obtained in our experiment are capable of visualizing convective flows from the
shadow-graph effect. We have not seen any evidence of the steady convection that is required to create and
maintain the observed bubble shape continuously during the heating. We conclude that the Marangoni convection
is absent.

This conclusion is an apparent contradiction with many works that study the Marangoni effect caused by
evaporation (see e.g. \cite{Palmer}). The main difference between these works and ours is in the conditions of
evaporation. These works consider the evaporation into an \emph{open} space where the vapor pressure is very
small. The interface temperature thus follows the temperature in the bulk of the liquid and a very large
evaporation rate is possible, limited only by the average velocity of the fluid molecules. In our case, the
gas phase is almost at saturation pressure. This means that the total evaporation (over
 the whole gas liquid interface) is small and limited by the amount of the supplied heat consumed by the
latent heat. Therefore, any variation $\delta T_i$ is rapidly dampened by the corresponding change in the
evaporation rate, stabilizing the interface against Marangoni convection, see \cite{Straub} for an extended
discussion. This conclusion is confirmed by the experiments \cite{Barnes}, in which Marangoni convection was
carefully studied in a \emph{closed} cell with very clean water in contact with its vapor. \emph{No}
surface-tension-driven convection was registered in spite of a large Marangoni number that was much greater
than its critical value obtained in the classical Marangoni-Benard experiments with non-volatile liquids
\cite{7}. It was argued in \cite{7} that the convection was absent due to a hypothetical interface
contamination present in spite of many careful preventive measures. According to our reasoning, a variation
$\delta T_i$ would have been strongly dampened in \cite{7} because of the saturation conditions in the sealed
cell. We also note that even in evaporative driven Marangoni convection far from saturation, the convection
cells may also tend to stabilize the interface resulting in intermittent cellular formation as was observed
in \cite{8}. It was also observed in \cite{8} that the velocity of convection and frequency of intermittent
cell formation decreases as the external gas becomes more saturated.

\subsection{Differential vapor recoil}

We now analyze another possible source of bubble deforming stress that does not
require a temperature
gradient along the interface. The bubble may be deformed by the normal stress exerted on the interface by the
recoil from departing vapor \cite{Palmer}. Let $n(\vec{x})$ be the evaporating mass per unit time per unit
interface area at the point $\vec{x}$ on the interface. The evaporating gas moves normally to the interface,
and exerts a force per unit area (a ``thrust") on the liquid of

\begin{equation}\label{Pr}
  P_r(\vec{x})=n^{2}(\vec{x})(1/\rho_G-1/\rho_L),
\end{equation}
where $\rho$ denotes mass density and the subscripts $L$ and $G$ refer to liquid
and gas respectively.

The interface shape can be obtained from a quasi-static argument when the experimentally observed interface
velocity $v_i$ is smaller than the characteristic hydrodynamic velocity $\sigma/ \eta$, where $\eta$ is the
shear viscosity. A numerical estimate shows that the quasi-static approximation holds for the images (a-f) in
Fig.~\ref{GMSF2}, in which the spreading is observed. The quasi-static approximation does not appear to hold
for the quench experiments (Fig.~\ref{Pegase4608}), where the interface moves rapidly.

According to the quasi-static argument \cite{13}, the interface shape can be determined from the modified
Laplace equation
\begin{equation}\label{modLap}
\sigma K=\Delta p+P_r(\vec{x}).
\end{equation}
The 3D curvature $K$ is equal to the sum of the 2D curvature $c$ in the image plane and the 2D curvature in
the perpendicular plane shown in Fig.~\ref{fig1}. For the small cell thickness $H$, this latter curvature can
be accurately approximated by the constant value $2/H$. This is possible because the relatively small heat
flow through the less conductive sapphire windows implies a small $P_r$ near the contact line on the windows,
as compared to the large value of $\Delta p$ at this small $H$. The interface shape can thus be obtained from
the 2D equation
\begin{equation}\label{modLap1}
\sigma c=\Delta p'+P_r(l),
\end{equation}
where $\Delta p'$ is a constant to be determined from the known bubble volume
and $l$ is a coordinate that varies along the bubble contour in the image
plane.

In order to find the distribution $n(\vec{x})$ at the interface it is necessary to solve the entire heat
transfer problem. This problem is complicated by several important factors. First, we deal with a problem
that contains a free boundary (gas-liquid interface) the position of which should be determined. Second, this
interface contains lines of singularities (gas-liquid-solid contact lines) where various divergences are
possible. Third, the adiabatic heat transfer \cite{Straub,11, Reg} (``the piston effect") should be taken
into account for near-critical fluids. The first two complications were addressed in \cite{13,IJMF,IJHMT} for
plane geometry, i.e. for the gas bubble growing on a plane. We have shown that $n(\vec{x})$ can exhibit a
divergence at the contact line and that it decreases exponentially far away from it. Because the bulk
temperature varies sharply in the boundary layer adjacent to the walls of the cell \cite{11} and the interface
temperature is constant, the largest portion of mass transfer across the interface takes place near the
triple contact line. Thus $n(\vec{x})$ is large in the vicinity of the contact line. In this work, we present
first the scaling arguments and then an approximate calculation of the bubble shape to illustrate our
explanation of the spreading gas in the cylindrical geometry.

We assume that $n(\vec{x})$ has the following form:
\begin{equation}\label{nsc}
  n(\vec{x})=g(\vec{x})(T_c-T)^{a}
\end{equation}
as $T\rightarrow T_c$, i.e., it has the same local behavior with respect to temperature as the critical
temperature is approached. The integral rate of change of mass $M$ of the gas bubble is defined as
\begin{equation}\label{M1}
  {\rm d}M/{\rm d}t=\int n(\vec{x}){\rm d}\vec{x}\sim(T_c-T)^{a},
\end{equation}
where the integration is performed over the total gas-liquid interface area. On
the other hand,
\begin{equation}\label{M2}
  {\rm d}M/{\rm d}t={\rm d}/{\rm d}t(V\phi\rho_{G}),
\end{equation}
where $V$ is the cell volume, and $\phi =0.5$ is assumed. Near the critical point, the co-existence curve has
the form $\rho_G=\rho_c-\Delta\rho/2$, where $\Delta\rho\sim(T_c-T)^\beta$ with the universal exponent
$\beta=0.325$, so that ${\rm d}M/{\rm d}t \sim(T_c-T)^{\beta -1}{\rm d} T/{\rm d}t $ as $T\rightarrow T_c$
according to Eq.~(\ref{M2}). Thus Eq.~(\ref{M1}) results in $a=\beta -1$ and the curvature change due to the
vapor recoil scales as
\begin{equation}\label{prs}
  P_r /\sigma\sim(T_c-T)^{3\beta-2-2\nu},
\end{equation}
where Eq.~(\ref{Pr}) and the scaling relationship $\sigma\sim(T_c-T)^{2\nu}$ ($\nu=0.63$) were employed.
Because this critical exponent $(3\beta-2-2\nu\approx -2.3)$ is very large, it should manifest itself even
far from the critical point in agreement with the experiments. In summary, as $T\rightarrow T_c$, the vapor
mass growth follows the growth of its density (the vapor volume remains constant), so that the diverging vapor
production near the critical point drives a diverging recoil force.

This curvature change has a striking effect on the bubble shape because it is not homogeneously distributed
along the bubble interface. Since the evaporation is strongest near the copper heating wall where the
strongest temperature gradients form, both $P_r$ and $c$ increase strongly near this wall, i.e. near the
triple contact line. Note that $c$ is proportional to the second derivative of the bubble shape function,
i.e. to the first derivative of the bubble slope. If $c$ is large, then the slope of the bubble contour
changes sharply when moving along the bubble contour towards the contact line, see \cite{IJT,13,IJMF} for
more details. Because the interface slope changes so abruptly near the contact line, the apparent contact
angle should be much larger than its actual value.

Because $c$ is proportional to the second derivative of the bubble shape function, Eq.~(\ref{modLap1}) is a
differential equation with the boundary condition given by the actual contact angle \cite{13}. This actual
contact angle defines the first derivative (the slope) of the bubble shape function at the solid wall. It is
also specified by the interfacial tension balance and must be zero near the critical point. This condition of
the zero contact angle gives a boundary condition for Eq.~(\ref{modLap1}). In order to illustrate a possible
solution of Eq.~(\ref{modLap1}), we solved it using the same expression for $P_r(l)$ as in \cite{13}
\begin{equation}\label{Pr-ex}
  P_r(l)\propto -N\,\log (l/L)\,\exp\{(-[l/(0.1\,L)]^2\},
\end{equation}
where $l\in[0, L]$, $L$ being a length of the bubble half-contour with $l=0$ at
the solid wall. We use a
non-dimensional parameter $N$ to measure the influence of the vapor recoil force
relative to the surface
tension. It is defined as
\begin{equation}
N={1\over\sigma}\int\limits_0^L P_r(l)\;{\rm d}l. \label{Nr}
\end{equation}
where the integration is performed over the drop contour in the image plane. The numerical coefficient (see
\cite{13}) in Eq.~(\ref{Pr-ex}) can be determined from  Eq.~(\ref{Nr}), where the upper integration limit can
be replaced by infinity without any loss of accuracy. Although the expression Eq.~(\ref {Pr-ex}) for the vapor
recoil pressure is not rigorous, it contains the main physical features of the solution of the heat conduction
problem: a weak divergence at the contact line and a rapid decay away from it. It is shown in \cite{IJHMT}
that the rigorous numerical solutions obtained far from the critical point follow this behavior.

\begin{figure}[htb]
  \begin{center}
  \includegraphics[height=6cm]{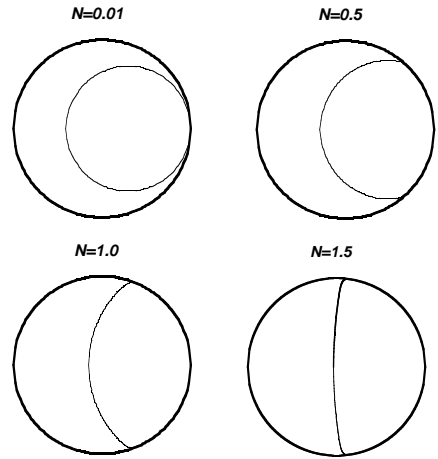}
  \end{center}
\caption{ Calculated bubble shape for different values of the non-dimensional
strength of vapor recoil $N$ that goes to infinity at the critical point. Note
that the actual contact angle is zero for all the curves.}\label{fig3}
\end{figure}

The result of this calculation is shown in Fig.~\ref{fig3}.
Since Eq.~(\ref{prs}) implies
\begin{equation}\label{Ns}
  N\sim(T_c-T)^{-2.3}\rightarrow\infty
\end{equation}
as $T\rightarrow T_c$, the $N$ increase mimics the approach to the critical point and qualitatively explains
the observed shape of the vapor bubble (see Fig.~\ref{GMSF2}). The increase of the apparent contact angle and
of the gas-solid contact area $A_{cg}$ can be seen in Fig.~\ref{fig3}. Note that such a calculation is not
able to predict the wavy interface shapes like those in Fig.~\ref{GMSF2}g or Fig.~\ref{GMSF3}f-h, because
these images correspond either to $T>T_c$ (images g and h of the both figures) or to the close vicinity of
$T_c$ where $\sigma<v_i\,\eta$, see the discussion of the validity of the quasi-static approximation earlier
in this section.

\section{Spreading gas and the boiling crisis}\label{sec-b}

A very similar bubble spreading was observed far from $T_c$ during boiling at large heat flux \cite{Van,Tor}.
When the heating to a surface is increased past a Critical Heat Flux (CHF) there is a sudden transition to
``film" boiling, where the heater becomes covered with gas and may burnout \cite{Tong}. This ``boiling
crisis" is an important practical problem in many industries where large heat fluxes are frequently used. We
interpret \cite{13,IJHMT} the boiling crisis to be similar to the gas spreading shown here. The main
difference is that the large value of $N$ is made by a large vapor production that can be achieved during
strong overheating rather than by the critical effects.

It is well-documented from experiments \cite{Tong} that the CHF decreases rapidly when the fluid pressure $p$
approaches the critical pressure $p_c$, i.e., when $T\rightarrow T_c$ in our constant volume system.
Previously, this tendency has not been well understood. The divergence of the factor $N$, discussed above,
helps to understand it. We first note that the evaporation rate $n$ scales as the applied heat flux $q$ and
$N\sim q^2$, where Eqs.~(\ref{Pr}) and (\ref{Nr}) are used. By assuming that the boiling crisis ($q=q_{CHF}$)
begins when $N$ attains its critical value $N_{CHF}\sim 1$ (see \cite{13}), one finds that
\begin{equation}\label{tchf}
  q_{CHF}\sim(T_c-T)^{1+\nu-3\beta/2}\sim(T_c-T)^{1.1}
\end{equation}
from Eq.~(\ref{Ns}). The same exponent is also valid for the pressure scaling,
\begin{equation}\label{pchf}
  q_{CHF}\sim(p_c-p)^{1.1}.
\end{equation}
Eq. (\ref{pchf}) explains the observed tendency $q_{CHF}\rightarrow 0$ as $p\rightarrow p_c$.

Although the strict requirements on temperature stability and the necessity of weightlessness lead to
experimental difficulties to study the boiling crisis in the near-critical region, they also present some
important advantages. Only a very small heating rate (heat flux) is needed to reach the boiling crisis because
$q_{CHF}$ is very small. At such low heat fluxes, the bubble growth is extremely slow due to the critical
slowing-down. In our experiments we were able to observe the spreading gas (i.e. the dry-out that leads to
the boiling crisis, see Fig.~\ref{GMSF2}) during 45~min! Such experiments not only permit an excellent time
resolution, but also allow the complicating effects of rapid fluid motion to be avoided.

\section{Conclusions}

In our experiments we observed a gas bubble spreading over a solid wall and a large value ($>90^\circ$) of the
apparent contact angle that appeared despite the zero actual contact angle with the solid. The spreading gas
is a phenomenon that can occur in a sealed heated fluid cell only when the bubble is pressed against the
heating wall. The 3D numerical calculation of the equilibrium bubble shape showed that the slightly tilted
windows of the experimental cell pressed the bubble against the copper side-wall. Weightless conditions are
needed in the near-critical region in order to observe this phenomenon when the surface tension is small and a
bubble-like shape persists. The same phenomenon can be observed far from the critical point during boiling at
high heat fluxes where it is known as the ``boiling crisis". While the gas spreads very quickly during the
boiling crisis far from the critical point, the near-critical region allows a very slow spreading gas to be
observed in great detail.

We explain this phenomenon as induced by the vapor recoil force that changes the shape of the vapor-liquid
interface near the triple contact line. Our preliminary calculations of the gas-liquid interface shape are
qualitatively consistent with the experimental images. The scaling analysis gives the critical exponent for
the critical heat flux decrease near the critical point and explains the increase of the vapor recoil effect
near the critical point. We believe that there is much to be learned about the boiling crisis in the
near-critical region and hope that these experiments inspire more investigations.

\begin{acknowledgments}
This work was supported by CNES and NASA Grant NAG3-1906. A part of this work was made during the stay of V.
N. at the UNO and he would like to thank the Department of Physics of the UNO for their hospitality. We thank
all of the Alice II team and everyone involved in the Mir missions. We especially thank J. F. Zwilling, and
the french cosmonauts Claudie Andr\'e-Deshays, L\'eopold Eyharts, and Jean-Pierre Haigner\'e. We thank Kenneth
Brakke for creating the Surface Evolver software and for making it available for the scientific community.
\end{acknowledgments}

\end{document}